\documentclass[12pt]{article}

\usepackage[T1]{fontenc}
\usepackage{ae,aecompl}

\usepackage[dvips]{graphicx}
\usepackage[usenames,dvipsnames]{color}
\usepackage{longtable}  
\usepackage{amssymb}
\usepackage{longtable}
\usepackage{rotating}
\usepackage{multirow}
\usepackage{float}
\usepackage{url}

\usepackage[T1]{fontenc}
\usepackage{calc}
\usepackage{setspace}
\usepackage{color}

\begin{document}

\title{Initial Experiences Re-exporting Duplicate and Similarity Computations with an OAI-PMH Aggregator}
\author {
	Terry L. Harrison, Aravind Elango, Johan Bollen, Michael L. Nelson\\
	Computer Science Department\\
	Old Dominion University\\
	Norfolk VA 23529\\
	\{tharriso,aelango,jbollen,mln\}@cs.odu.edu	
	}

\maketitle

\abstract{
The proliferation of the Open Archive Initiative Protocol for Metadata Harvesting (OAI-PMH) has resulted in
the creation of a large number of service providers, all harvesting from either data providers or aggregators.
If data were available regarding the similarity of metadata records, service providers could track 
redundant records across harvests from multiple sources as well as provide
additional end-user services.  Due to the large number of metadata formats and the diverse
mapping strategies employed by data providers, similarity calculation requirements necessitate the use 
of information retrieval strategies.  We describe an OAI-PMH 
aggregator implementation that uses the optional ``<about>'' container to re-export the results of 
similarity calculations. Metadata records (3751) were harvested from a NASA data provider and similarities
for the records were computed. The results were useful for detecting duplicates, similarities and metadata
errors. }

\section{Introduction}
\subsection{Problem Statement}
The Open Archive Initiative Protocol for Metadata Harvesting (OAI-PMH) is a succinct, six verb protocol
for the dissemination of metadata and represents the de-facto metadata exchange standard for today's digital
libraries \cite{oai:Sompel}. Yet with over six million publicly available records, no general purpose tools
exist to aid service providers in discovery of similarity among harvested records. \\

 \begin{figure}
       \begin{center}
       \includegraphics[scale=0.5]{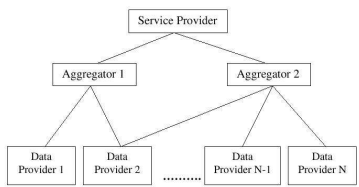}
       \end{center}
       \caption{\label{duplicate_case1} A possible scenario where duplicate records may be harvested via separate aggregators.}
 \end{figure}

\begin{figure}
       \begin{center}
       \includegraphics[scale=0.5]{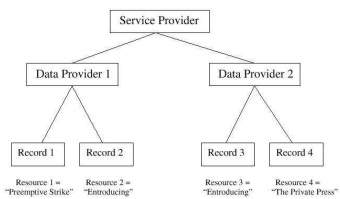}
       \end{center}
       \caption{\label{duplicate_case2} A possible scenario where duplicate records may be harvested through multiple descriptions of the same resource.}
 \end{figure}

 OAI-PMH increases the venerable problem of duplicates in union catalogues
.
 Figures \ref{duplicate_case1} and \ref{duplicate_case2} depict scenarios where similar or duplicate
 records could be harvested. Figure \ref{duplicate_case1} depicts a scenario where a service provider 
 has twice harvested from Data Provider 2. Perhaps Aggregator 1 or Aggregator 2 have performed some metadata
 normalization, resulting in new OAI identifiers being assigned.
 ``Sameness'' becomes harder to compute. The <provenance> records can help identify duplicates, but they are optional.
 Figure \ref{duplicate_case2} shows a service provider harvesting from two separate data providers. 
Both data providers have reviews of DJ Shadow LPs. Are the two reviews of ``Entroducing'' the same, 
or are they merely similar ? \\

\subsection{Proposed Solution}

A general purpose system devised to measure the similarity of OAI metadata records using information
retrieval (IR) methodologies has many valuable uses.  It may be used to weed out duplicate records that might
otherwise be difficult to find by more traditional field matching methodologies. Also, it could be
used to find additional versions of the same work, such as locating the short and long paper versions 
of the same project.  Finally, this system could find similar documents in accordance with a predetermined
threshold for use in recommendation systems such as those described in \cite{irNotes:johan2003}.\\

Such a ``similarity engine'' has been built to calculate similarity among 
harvested metadata records and thereby detect duplicate or similar documents. Our OAI-PMH aggregator incorporates these results, appending 
similarity data to metadata record requests. When issued a ``GetRecord'' request, the standard metadata for 
the identifier is returned along with a ranked list of similar documents. Ranking values are between 0 
and 1, where 1 represents absolute similarity. This work is intended as a proof of concept and will require 
further optimization before dealing with the open corpus of OAI-PMH metadata (over 6.5 million records).\\

The Vector Space Model (VSM) \cite{termWeigh:1988} computes the cosine similarity of documents based on the common terms present in the \
documents and their approximate importance. The importance of a term is gauged by its frequency of occurrence 
within a document and its sparse presence in the rest of the document collection. We use VSM as the
primary mechanism for computing the similarity among the metadata.  Salton describes the
formula used as the best fully weighted system and recommends its use in the processing of text abstracts \cite{termWeigh:1988},
which is similar to the text found in a metadata record. 
Our aggregator returns the identifier similarities within the context of the standard OAI-PMH 
``GetRecord'' request. A service provider could use this data to further process its records (e.g. creating 
links to similar documents, deleting them, logging them for examination, etc.).\\

\section{Methodology}
\begin{figure}[H]
       \begin{center}
       \includegraphics[scale=1.0]{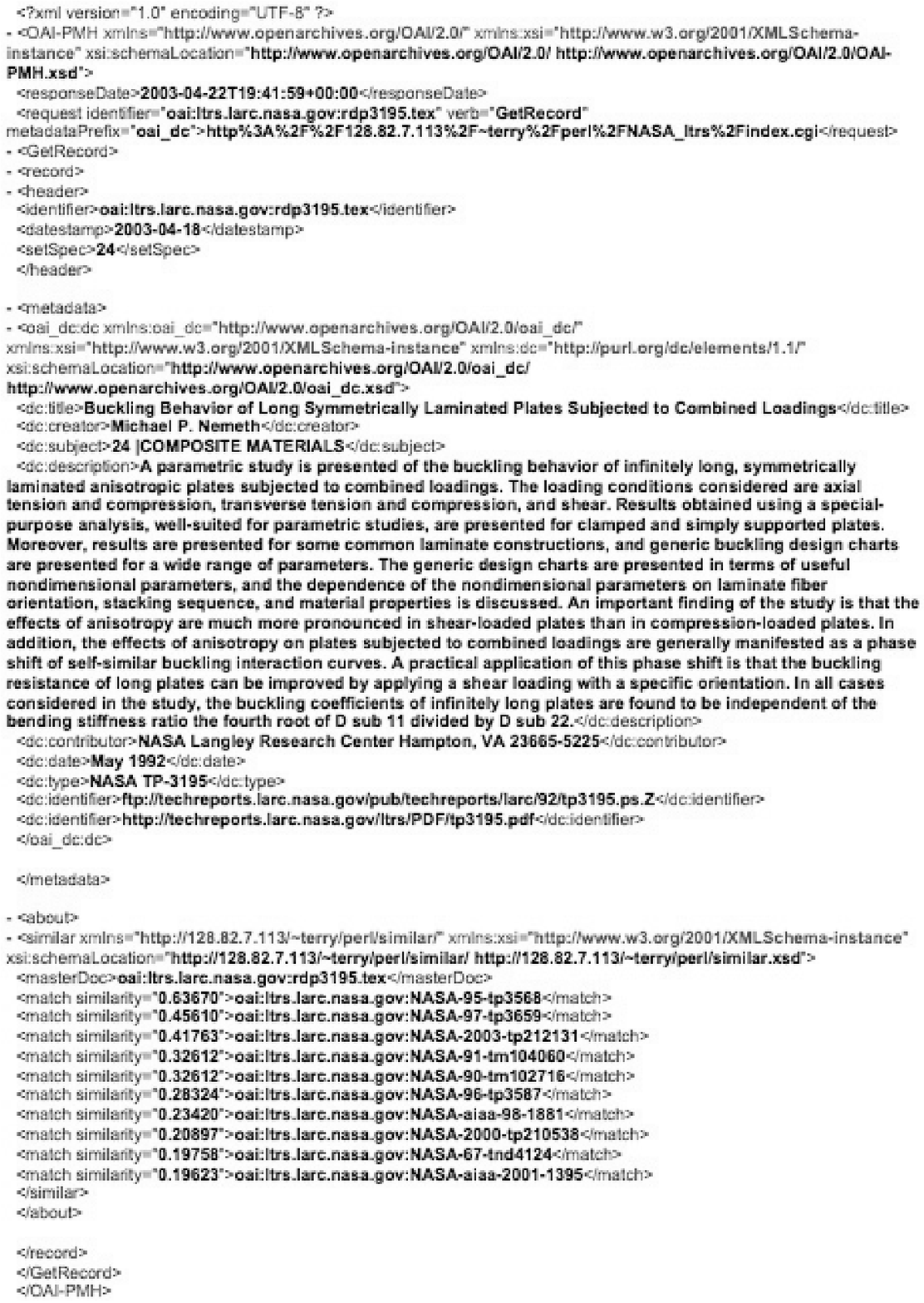}
       \end{center}
       \caption{\label{record} Sample GetRecord response matching identifier.}
 \end{figure}

For proof of concept purposes, NASA's Langley Technical Reports Server (LTRS) \cite{ntrs:nelson2003} served as the repository to 
generate our test collection of 3751 records. To keep interface complexities to a minimum, only a 
predetermined (ten) number of matches (top similar documents and their similarity score) are returned and 
appended to the GetRecord response. Currently this number is set by our aggregator, but could easily 
be passed by the client as well. The following URL could yield the result shown in Figure \ref{record}.\\\\

\url{http://128.82.7.113:5180/perl/NASA_ltrs/?verb=GetRecord&metadataPrefix=oai_dc&identifier=oai:ltrs.larc.nasa.gov:rdp3195.tex}\\\\

Use of the VSM permits the makeup of the record collection to affect the importance of each term
in a given document. Given the large number of OAI-PMH records that would ultimately be harvested and
evaluated, it would be extraordinarily costly to compare the full text documents. Instead, the metadata
records are  used. Individual service providers could be relieved of the expensive burden of calculating record 
similarity and may instead query (and even re-harvest) this data computed by the similarity 
engine.\\

Harvested records are cached in a hierarchical file structure. This structure is duplicated to store 
each file's term list/frequencies (in a tf\_metadata directory) and again to store each file's term weights (in 
weights\_metadata) after idf has been calculated. Since additional harvests will add additional records, this will 
require recalculating collection idf weights and similarities. Note that idf values are 
not cached as these values are calculated at runtime. Once the tf\_metadata directory is created, it could be 
used for subsequent collection calculations. This does not hold for the weight\_metadata files, for their 
values are influenced by the runtime idf weight calculations.\\

\section{Results}

LTRS was harvested in  the first week of April 2003, yielding 3751 records. The similarity engine
calculated the similarities of these 3751 records in just over 7 hours. This is taking into account document term parsing, document term
frequency calculation, idf, term weights for each document and the similarity calculations per document  into the total runtime. Dividing
this by the number of documents we can estimate the cost of a single document to document comparison at approximately .0036
seconds.\\

Given this cost per similarity calculation, we can get an idea of how this would scale.  
To compare a document's similarity to all other documents  requires an order O($n^2$) operation, even if 
we are only calculating the upper triangular portion of a document to document matrix O($((n^2 - n)/2))$.
Reducing this order of complexity is beyond the scope of this proof of concept, although we are currently 
investigating other techniques. The estimated similarity computation time for an increasing number of records
is shown in table \ref{est_graph}.\\

 \begin{table}
 \begin{center}
 \begin{tabular}{|c||r|}
 \hline
\textbf{Number of Records}		&	\textbf{Estimated Computation Time} \\ \hline & \\
 100	&  17 seconds	\\ 
 1000	&  30 minutes-1 second	\\ 
 5000	&  12 hours-31 minutes-18 seconds    \\
10000	&  2 days-2 hours-5 minutes-30 seconds	 \\
25000	&  13 days-1 hour-5 minutes-31 seconds	 \\
50000	&  52 days-4 hours-23 minutes-36 seconds \\ 
100000	&  208 days-17 hours-37 minutes-24 seconds \\
1000000	&  57 years-68 days-14 hours-51 minutes-36 seconds \\
6500000	&  2416 years-71 days-3 hours-45 minutes-2 seconds	\\ & \\
 \hline
 \end{tabular}
 \caption[short title here]{\label{est_graph} Estimated similarity computation time for an increasing number of records.}
 \end{center}
 \end{table}

The computed results were written to the file ``similarities.txt''.  
The top ten collection matches and their nature of similarity are represented in (Appendix 2). To further
process the results, a script was written which read ``similarities.txt'' and created a directory of results. 
Each file in it was named after a document and represented the top 10 (user definable) closest matches found,
sorted high to low (table \ref{topTenFiles}). With this built, the results were ready to be exported.\\

 \begin{table}
       \begin{center}
       \includegraphics[scale=.6]{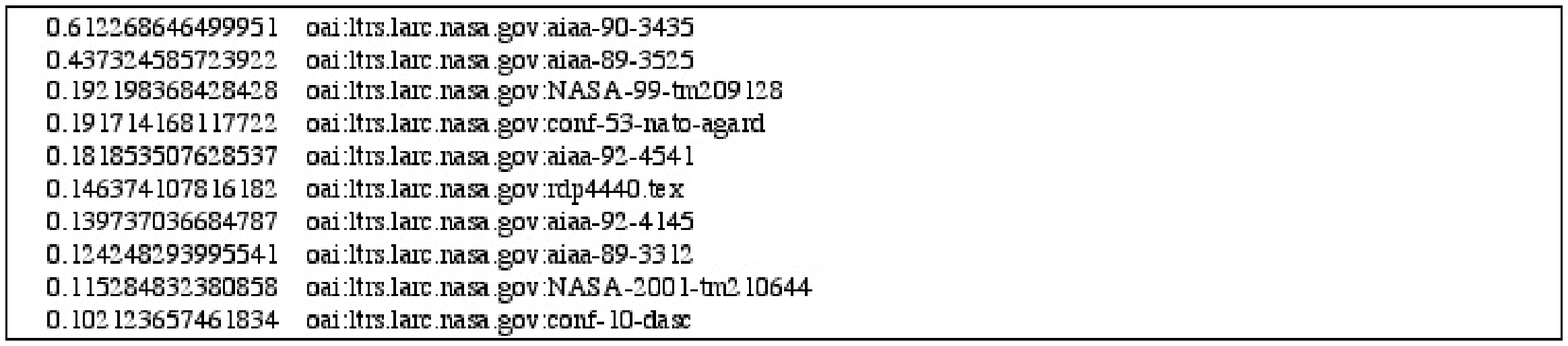}
       \end{center}
       \caption{\label{topTenFiles} Top ten matches in LTRS collection to file oai:ltrs.larc.nasa.gov:9-dasc. }
 \end{table}

A fully functional OAI-PMH compliant aggregator was built that used the harvested records as its repository.  Added to the
data provider was the ability for it to serve the similarity information along with the other metadata normally provided. This
modification was done within the scope of the OAI-PMH, thus maintaining OAI-PMH compliance. The similarity data is housed within an
<about> section, which is available as an optional part of a GetRecord response. An XML schema was written for this container, as per
OAI-PMH specifications (Appendix 1). \\

The average similarity calculation time between every two documents took approximately 0.0036 seconds.
The result of 7,033,125 such similarities was stored in a 678.1 MB file (98.7 MB if compressed).
The disk space for the harvested LTRS collection, tf\_metadata and weight\_metadata were 15.2 MB, 
15.2 MB and 17.3 MB respectively.

\section{Conclusions}
We have demonstrated a proof of concept showing the use of information retrieval technologies
with the OAI-PMH. This cross fertilization yields valuable results. At the data provider, similarity results
may be used in detection of duplicates within the collection as well of the location of related documents (e.g. the Vol 1 / Vol 2 scenario).
This information can be used for further grouping and association making. Through the interface demonstrated the user is provided with
top ten (in our case) matches. A harvester could use this information to create links to these associated documents for the end
user to peruse. Such links could direct users to alternative versions of documentation or to subsequent parts of the same
report.\\

This project is only the beginning of a greater investigation into the similarity evaluation of OAI-PMH metadata records. The
algorithm used here is O($n^2$), which poses great scaling issues. While there are many optimizations which may still be made, including
parallelizing the calculations through a distributed calculation mechanism, in the end another algorithm must
be employed.\\

\bibliographystyle{plain}
\bibliography{project}

\section*{Appendix 1. XML schema for similarity in <about> container}

 \begin{figure}[H]
       \begin{center}
       \includegraphics[scale=.6]{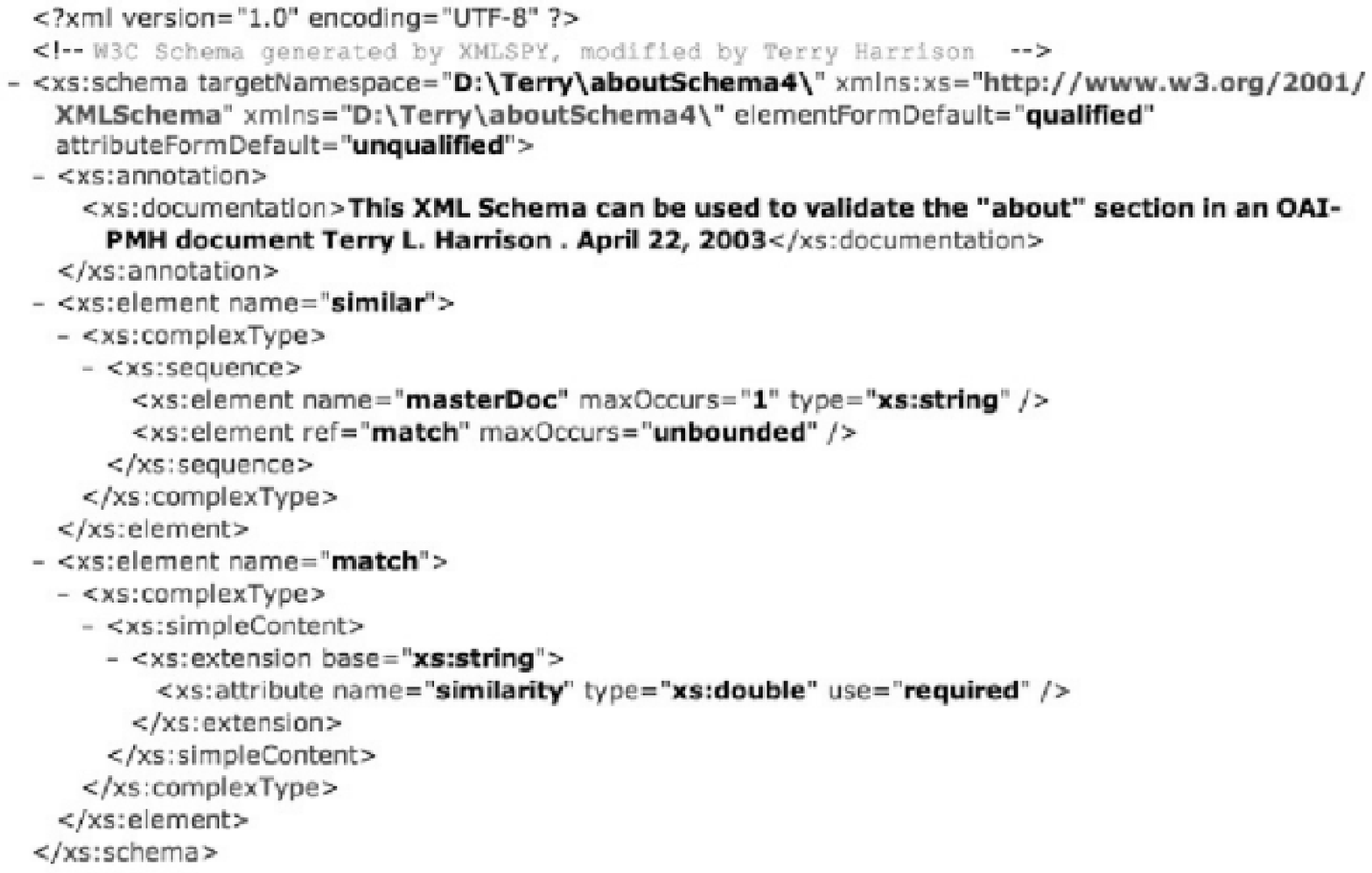}
       \end{center}
 \end{figure}

\newpage

\section*{Appendix 2. Top 10 similar document pairs in LTRS collection}
        \begin{footnotesize}
        \begin{longtable}{|p{1cm}|p{7.5cm}|p{1.5cm}|p{4.5cm}|}
        \hline\hline
        \textbf{Rank} & \textbf{Titles of Similar Documents}  & \textbf{Similarity} & \textbf{Relatedness}\\\hline
        &  & & \\
        
	1 & Space Environmental Effects on Spacecraft:LEO Materials Selection Guide (oai:ltrs.larc.nasa.gov:NASA-95-cr466lpt1)	& 0.9822 & Part 1 and Part 2 of same report. \\       
        & Space Environmental Effects on Spacecraft:LEO Materials Selection Guide (oai:ltrs.larc.nasa.gov:NASA-95-cr466lpt2) &  & \\
        & & & \\
        
	2 & Nondimensional parameters and equations for Buckling of Symmetrically Laminated Thin Elastic Shallow Shells (oai:ltrs.larc.nasa.gov:NASA-90-tm102716) & 0.980 & Suspected metadata error. \\
        & Nondimensional parameters and equations for Buckling of Symmetrically Laminated Thin Elastic Shallow Shells (oai:ltrs.larc.nasa.gov:NASA-91-tm104060) &  & \\
        & & & \\

	3 & Development of Pneumatic Channel Wing Powered-Lift Advanced Super-STOL Aircraft (oai:ltrs.larc.nasa.gov:NASA-aiaa-2002-2929) & 0.9617   & Same paper (or close match) submitted to two conferences having different identifiers. \\ 
	  & Pneumatic Channel Wing Powered-Lift Advanced Super-STOL Aircraft (oai:ltrs.larc.nasa.gov:NASA-aiaa-2002-3275)   &  & \\
        & & & \\

	4 & Compendium of NASA Data Base for the Global Tropospheric Experiment's Pacific Exploratory Mission-Topics B(PEM-Tropics B)---Volume1:DC-8. (oai:ltrs.larc.nasa.gov:NASA-2000-tm210617vol1)	 & 0.958 & Volume One and Volume Two of same report. \\
	  & Compendium of NASA Data Base for the Global Tropospheric Experiment's Pacific Exploratory Mission-Topics B(PEM-Tropics B)---Volume2:P-38. (oai:ltrs.larc.nasa.gov:NASA-2000-tm210617vol2)	& & \\
        & & & \\

	5 & Computational Methods for Frictional Contact With Applications to the Space Shuttle Orbiter Nose-Gear Tire-Development of Frictional Contact Algorithm.(oai:ltrs.larc.nasa.gov:NASA-96-tp3573)   & 0.956   & Suspected metadata error. \\
 	  & Computational Methods for Frictional Contact With Applications to the Space Shuttle Orbiter Nose-Gear Tire-Development of Frictional Contact Algorithm.(oai:ltrs.larc.nasa.gov:NASA-96-tp3574)	  &  &  \\ & & & \\\hline  

	6 & A Cryogenic Magnetostrictive Actuator Using a Persistent High Temperature Superconducting Magnet, Part 1: Concept and Design.(oai:ltrs.larc.nasa.gov:NASA-2000-tm209139)  &  0.950  & Technical Report version of a conference paper. \\ 
	  & A Cryogenic Magnetostrictive Actuator Using a Persistent High Temperature Superconducting Magnet, Part 1: Concept and Design.(oai:ltrs.larc.nasa.gov:NASA-99-6spie-gch1)			   & & \\
        & & & \\
	7 & International Space Station Evolution Data Book-Volume1.(oai:ltrs.larc.nasa.gov:NASA-2000-sp6109vol1rev1)	  &  0.937  &  Volume One and Volume Two of same report. \\ 
	  & International Space Station Evolution Data Book- Volume2 Evolution Concepts-Revision A.(oai:ltrs.larc.nasa.gov:NASA-2000-sp6109vol2rev1)  &  &  \\ 
        & & & \\
	8 & A Model for Assessing the Liability of Seemingly Correct Software.(oai:ltrs.larc.nasa.gov:NASA-96-icai.jlr)  & 0.935  & Technical Report version of a conference paper.  \\ 
	  & A Model for Assessing the Liability of Seemingly Correct Software.(oai:ltrs.larc.nasa.gov:NASA-96-tm110247)	  &  &	\\ 
        & & & \\
	9 & NASA's Atmospheric Effects of Aviation Project--- Result of the August 1999 Aerosol Measurement Intercomparison Workshop, Laboratory Phase.(oai:ltrs.larc.nasa.gov:NASA-96-iasted.jmv)	& 0.935  & Different phases of same project. \\
	  & NASA's Atmospheric Effects of Aviation Project--- Result of the August 1999 Aerosol Measurement Intercomparison Workshop, T-38 Aircraft Sampling Phase.(oai:ltrs.larc.nasa.gov:NASA-96-icrqcr-jmv)  &  & \\ 
        & & & \\
	10 & Integrated Orbit, Attitude and Structural Control Systems Design for Space Solar Power Satellites.(oai:ltrs.larc.nasa.gov:NASA-2001-tm210829) &	0.928  & Suspected metadata error. \\
	   & Integrated Orbit, Attitude and Structural Control Systems Design for Space Solar Power Satellites.(oai:ltrs.larc.nasa.gov:NASA-2001-tm111226) &  &	\\ 
        &  &  & \\\hline
        \end{longtable}
        \end{footnotesize}

\end{document}